%% file: main.tex
\documentclass[conference,a4paper]{IEEEtran} 
\IEEEoverridecommandlockouts

\usepackage{cite}
\usepackage{amsmath,amssymb,amsfonts}
\usepackage{graphicx}
\usepackage{textcomp}
\usepackage{xcolor}
\usepackage{acro}
\usepackage{url}
\usepackage{siunitx}
\usepackage{booktabs}
\usepackage{tabularx}
\usepackage{multirow}
\usepackage{todonotes}
\usepackage{algorithmicx}
\usepackage{algorithm}
\usepackage[noend]{algpseudocode}
\usepackage{mathtools}
\usepackage{graphicx}
\usepackage{subcaption}

\usepackage{lipsum}
\usepackage{makecell}

\usepackage{scalerel}
\usepackage{tikz}
\usepackage{pgfplots}
\pgfplotsset{compat=1.18}
\usetikzlibrary{svg.path}
\definecolor{orcidlogocol}{HTML}{A6CE39}
\tikzset{
  orcidlogo/.pic={
    \fill[orcidlogocol] svg{M256,128c0,70.7-57.3,128-128,128C57.3,256,0,198.7,0,128C0,57.3,57.3,0,128,0C198.7,0,256,57.3,256,128z};
    \fill[white] svg{M86.3,186.2H70.9V79.1h15.4v48.4V186.2z}
                 svg{M108.9,79.1h41.6c39.6,0,57,28.3,57,53.6c0,27.5-21.5,53.6-56.8,53.6h-41.8V79.1z M124.3,172.4h24.5c34.9,0,42.9-26.5,42.9-39.7c0-21.5-13.7-39.7-43.7-39.7h-23.7V172.4z}
                 svg{M88.7,56.8c0,5.5-4.5,10.1-10.1,10.1c-5.6,0-10.1-4.6-10.1-10.1c0-5.6,4.5-10.1,10.1-10.1C84.2,46.7,88.7,51.3,88.7,56.8z};
  }
}
\newcommand\orcidicon[1]{\href{https://orcid.org/#1}{%
\mbox{\begin{tikzpicture}[yscale=-1,transform shape, scale=0.03] 
\pic{orcidlogo};
\end{tikzpicture}}}}

\usepackage[english]{cleveref}
\usepackage{dsfont}
\usepackage{placeins}

\usepackage{xspace}
\let\OldTexttrademark\texttrademark
\renewcommand{\texttrademark}{\OldTexttrademark\xspace}%

\DeclareSIUnit{\kbps}{Kbit/s}
\DeclareSIUnit{\mbps}{Mbit/s}
\DeclareSIUnit{\gbps}{Gbit/s}
\DeclareSIUnit{\tbps}{Tbit/s}
\DeclareSIUnit{\B}{byte}
\DeclareSIUnit{\kB}{kbyte}
\DeclareSIUnit{\mB}{Mbyte}
\DeclareSIUnit{\gB}{Gbyte}
\DeclareSIUnit{\tB}{Tbyte}
\DeclareSIUnit{\b}{bit}
\DeclareSIUnit{\kb}{kbit}
\DeclareSIUnit{\mb}{Mbit}
\DeclareSIUnit{\gb}{Gbit}
\DeclareSIUnit{\tb}{Tbit}
\DeclareSIUnit{\mpps}{Mpps}

\newcolumntype{Y}{>{\centering\arraybackslash}X}

\def\BibTeX{{\rm B\kern-.05em{\sc i\kern-.025em b}\kern-.08em
    T\kern-.1667em\lower.7ex\hbox{E}\kern-.125emX}}

\input{macros.tex}
\input{acronyms.tex}

\usepackage{fancyhdr}
\fancypagestyle{IEEEtitlepagestyle}{
    \fancyhf{} 
     
    \fancyhead[C]{
        \small © 2026 IEEE. Personal use of this material is permitted. Permission from IEEE must be obtained for all other uses, including reprinting/republishing this material for advertising or promotional purposes, collecting new collected works for resale or redistribution to servers or lists, or reuse of any copyrighted component of this work in other works.\\
        \small \textit{Digital Object Identifier 10.1109/TBA}
        }
}
\begin{document}

\title{Performance Evaluation of Selection Strategies for Inter-Satellite Paths in Walker-Delta Constellations}

\author{
\IEEEauthorblockN{Marvin F. Braun, 
Moritz Flüchter,
and Michael Menth}
\IEEEauthorblockA{University of T\"ubingen, Chair of Communication Networks
\\Email: \{ma.braun,
moritz.fluechter, 
menth\}@uni-tuebingen.de}
  \thanks{This work was funded by ESA under contract no. 4000138133/22/UK/AL. The authors alone are responsible for the content.}
  }

\maketitle

\input{chapters/00-abstract}


\vspace{-0.3cm}

\input{chapters/01-introduction}

\input{chapters/02-related_work}

\input{chapters/03-system-model}
\input{chapters/04-metrics-selectors}
\input{chapters/05-evaluation}

\input{chapters/06-discussion}

\input{chapters/07-conclusion}

\bibliography{bibliography/literature}

\bibliographystyle{ieeetr}

\end{document}

%% file: macros.tex
\newcommand\citeN\cite

\DeclareMathOperator{\LA}{\delta^+_L}

\DeclareMathOperator*{\argmin}{arg\,min}
\DeclareMathOperator*{\argminlex}{arg\,min^{\mathrm{lex}}}
\DeclareMathOperator*{\argmax}{arg\,max}
\providecommand{\acsatu}{\alpha_u}
\providecommand{\acsatg}{\mathcal{A}_g}

\newcommand{\figepsNew}[3][]{%
   \begin{figure}[tb!]
        \centering
        \includegraphics[width=\columnwidth]{figures/#2}
         \caption{#3\vspace{-0.2cm}}
         \label{fig:#2}
      \vspace{-0.45cm}
   \end{figure}
}

\newcommand{\figepsFullwithvspace}[2]{
    \begin{figure*}[t]
    \vspace{2mm}
        \centering
        \includegraphics[width=\textwidth]{figures/#1}
        \caption{#2}
        \label{fig:#1}
    \end{figure*}
}

\newcommand{\figepsFullht}[2]{
    \begin{figure*}[ht]
        \centering
        \includegraphics[width=\textwidth]{figures/#1}
        \caption{#2}
        \label{fig:#1}
    \end{figure*}
}

\newboolean{makevspace}
\newcommand{\cvspace}[1]{%
   \ifthenelse
   {\boolean{makevspace}}
   {\vspace{#1}}
   {}%
}

%% file: acronyms.tex
\acsetup{single}

\DeclareAcronym{UT}{
    short = UT,
    long = user terminal,
    short-plural = UTs,
    plural = user terminals,
}

\DeclareAcronym{GW}{
    short = GW,
    long = gateway,
    short-plural = GWs,
    plural = gateways,
}

\DeclareAcronym{AcS}{
    short = AcS,
    long = access satellite,
    short-plural = AcSs,
    plural = access satellite,
}

\DeclareAcronym{VC}{
    short = VC,
    long = visibility cone,
    short-plural = VCs,
    plural = visibility cones,
}

\DeclareAcronym{ISL}{
    short = ISL,
    long = inter-satellite link,
    short-plural = ISLs,
    plural = inter-satellite links,
}

\DeclareAcronym{TG}{
    short = TG,
    long = traffic generator,
}

\DeclareAcronym{DDoS}{
    short = DDoS,
    long = distributed denial-of-service,
}

\DeclareAcronym{RTT}{
    short = RTT,
    long = round-trip time,
}

\DeclareAcronym{DuT}{
    short = DuT,
    long = device under test,
    short-plural = DuTs,
    plural = devices under test,
}

\DeclareAcronym{IAT}{
    short = IAT,
    long = inter-arrival time,
}

\DeclareAcronym{MAT}{
    short = MAT,
    long = match-action table,
}

\DeclareAcronym{CBR}{
    short = CBR,
    long = constant bit-rate,
}

\DeclareAcronym{TCAM}{
    short = TCAM,
    long = ternary content addressable memory,
}

\DeclareAcronym{byte}{
    short = B,
    long = byte,
}

\DeclareAcronym{NRMSD}{
    short = NRMSD,
    long = normalized root-mean-square deviation,
}

%% file: chapters/00-abstract.tex
\begin{abstract}
In LEO satellite constellations, traffic between a user terminal and a gateway is carried over a satellite path. As the satellite constellation rotates around Earth, a new path must be reselected repeatedly from a set of path candidates.

In this paper, we study the impact of path selection strategies on several metrics: path length in terms of Euclidean distance and hop count, path-change rate, and rate of used links.
These metrics are relevant because they affect either communication latency or the complexity of control and resource management.

We explain how path candidates are generated, define three heuristic path selection strategies, and evaluate them over a large set of UT--GW scenarios within a single shell of a Walker–Delta constellation with 1,156 satellites. Overall, the results show that path selection has a significant impact on both latency-related metrics and path churn.

\end{abstract}

%% file: chapters/01-introduction.tex
\section{Introduction}
\label{sec:introduction}

In low Earth orbit (LEO) satellite networks with inter-satellite links (ISLs), user terminals (UTs) and gateways (GWs) connect to one or more access satellites via feeder links, and their traffic is forwarded over an ISL path between the UT-side and GW-side access satellites. 
As the constellation rotates continuously around Earth, UTs and GWs frequently connect to new access satellites. 
\Cref{fig:exp_224_shortest_path.pdf}
illustrates an ISL path, together with the visibility cones and feeder links of UT and GW.

Selecting an ISL path between eligible access satellites, i.e., satellites available for UT-side or GW-side access, affects both Euclidean path length and hop count, both of which directly affect communication latency.
Since access satellites remain available for different periods of time, their selection also influences handover frequency.
Such handovers are critical operations that place load on the control plane and may cause packet delay or loss.
Therefore, fewer handovers are better.

Furthermore, the choice of access satellites and the ISL path between them influences how many different links are activated over time. Each newly used link may have to carry additional traffic, which increases the burden on resource management and may create overload situations. Consequently, using fewer links over time reduces both resource-management effort and the risk of transient overload.

\figepsNew{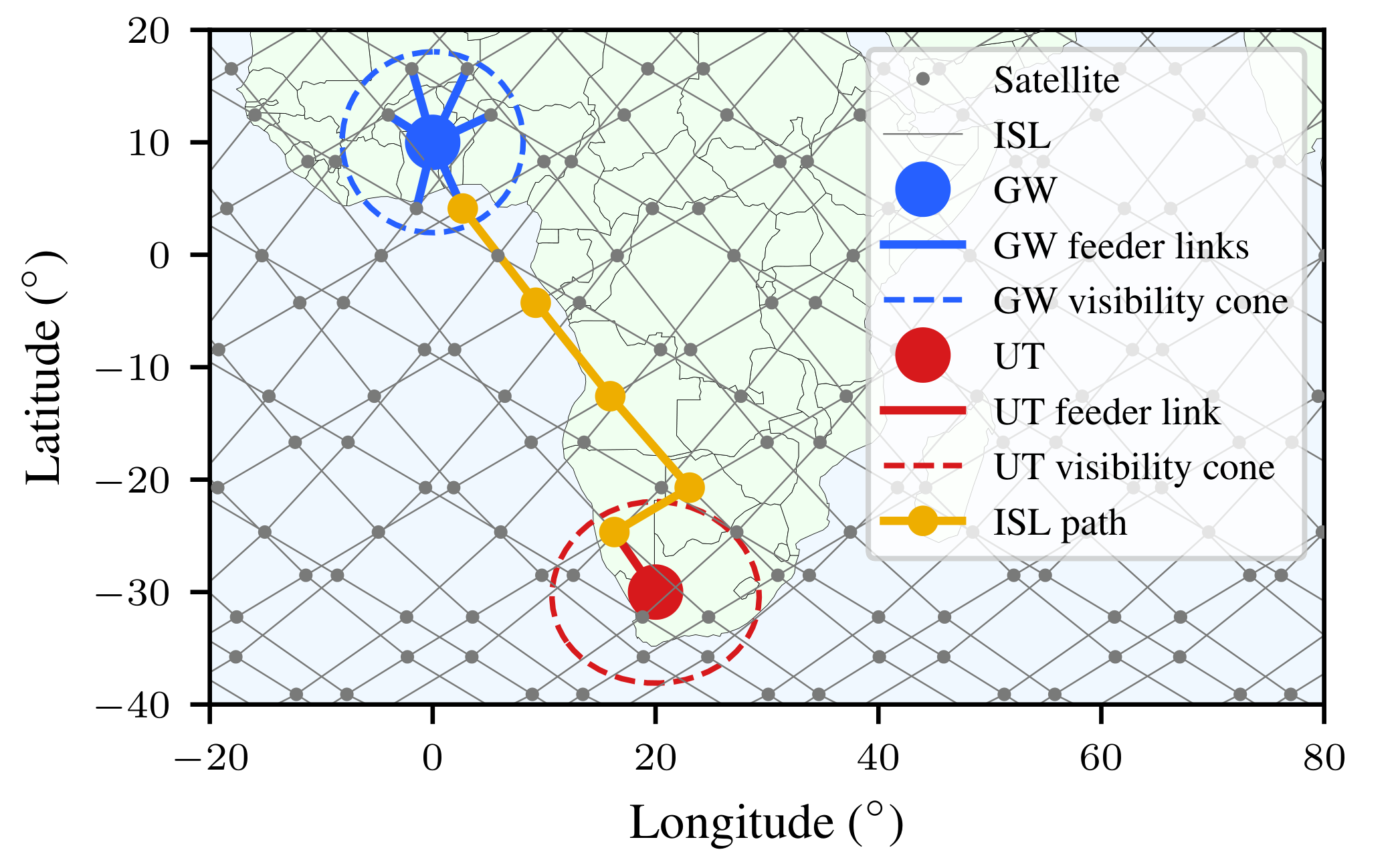}%
{UT--GW communication scenario that consists of a feeder link from the UT to a UT-side access satellite, an ISL path through the constellation, and a feeder link from one of several GW-side access satellites to the GW.}

In this work, we design different heuristic path selection strategies that choose an ISL path between access satellites such that they minimize the above mentioned metrics. 
We model a Walker–Delta LEO constellation consisting of 1,156 satellites on circular orbits at a common altitude and equal inclination. 
We consider various UT--GW combinations and simulate their handovers over time. 
The same set of eligible access satellites, determined by visibility duration, is used across all selectors to ensure comparability. 
We apply different selection strategies for choosing the ISL path between eligible access satellites, and compute the above mentioned metrics. 
The results demonstrate that path selection strategy has a significant impact on the considered metrics.

The remainder of this paper is organized as follows. 
\Cref{sec:rel_work} reviews related work. 
\Cref{sec:system-model} introduces the system model and nomenclature used throughout this paper.
\Cref{sec:metrics} formally defines the metrics and proposes for each of them a heuristic path selection strategy that should minimize that metric during operation.
\Cref{sec:eval} presents the evaluation methodology and \Cref{sec:results} discusses the simulation results.
\Cref{sec:conclusion} concludes the paper.

%% file: chapters/02-related_work.tex
\section{Related Work}
\label{sec:rel_work}
Prior work has examined several aspects of routing in LEO constellations, including scalable path computation, topological properties, and path stability.

\subsection{Path Computation in Dynamic LEO Networks}
A central line of research in LEO routing focuses on how shortest or near-shortest paths can be computed efficiently.
Stock et al.~\cite{StFr22} study scalable routing in large LEO mega-constellations and show that precomputing shortest routes for all satellite pairs over all time steps becomes impractical as constellation size grows. To address this, they propose distributed on-demand routing methods that approximate efficient routes at substantially lower computational cost than classical shortest-path algorithms such as Dijkstra.

Chen et al.~\cite{ChYa24} study shortest-path computation by exploiting the regular orbital structure and ISL distance patterns of LEO constellations, they derive the explicit phase-based algorithm STEPCLIMB.

\subsection{Routing Structure of 4-ISL Walker-Delta Constellations}
Chen et al.~\cite{ChGi21} analyze routing properties of Walker-Delta LEO mega-constellations under the 4-ISL topology, where each satellite is connected to two neighbors in its own orbital plane and two in adjacent planes. 

The resulting logical ISL network has a torus-like structure, because both the plane dimension and the in-plane dimension exhibit cyclic wrap-around. They further show that ascending/descending access satellite combinations induce different so called path modes with substantially different hop counts, so that the effective end-to-end path depends not solely on the ground locations but on the distance in the torus-like structure of the source-side and destination-side access satellites.

\subsection{Stability-Aware Routing and Path Churn}
Another line of work treats temporal instability as a problem in its own right.
Bhosale et al.~\cite{BhSa23} characterize route variability in LEO satellite networks and show that many reroutes yield only marginal latency gains, and that route variability harms higher-layer performance. Zhang and Yeung~\cite{ZhYe22} propose delay-bounded and delay-aware routing algorithms that explicitly reduce route changes while keeping latency increases small. Ron et al.~\cite{RoYu25} incorporate link churn
as an optimization objective. 

Zhou et al.~\cite{ZhLi23} model ground access through minimum-elevation constraints at gateways and user terminals. They show that frequent handovers on gateway feeder links cause interruptions, add routing overhead, and reduce overall backhaul quality; this is particularly critical because gateway regions concentrate traffic from many satellites.

Together, these studies demonstrate that stability is an important concern in dynamic LEO networking. However, they typically examine complete routing architectures, update mechanisms, or topology design. By contrast, our work takes a narrower view and isolates the path selection problem in a simplified setting: different path selectors are evaluated with respect to the same set of path candidates derived from the same underlying access satellite assignment.

%% file: chapters/03-system-model.tex
\section{System Model}
\label{sec:system-model}
We consider a Walker--Delta constellations in which all satellites orbit at a common altitude on circular orbits with equal inclination. 
\figepsFullwithvspace{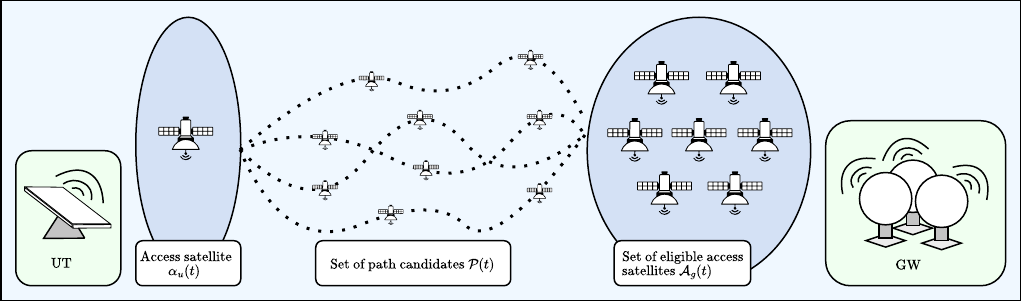}{%
Candidate-path construction. At time $t$, the user terminal is assigned one access satellite $\acsatu(t)$, the gateway is assigned a set of eligible access satellites $\acsatg(t)$. The set of candidate paths $\mathcal{P}(t)$ is formed by the union of the k-shortest paths from $\acsatu(t)$ to each eligible access satellite in $\acsatg(t)$. A path selector chooses one path from $\mathcal{P}(t)$.
}
To formalize the routing problem in a satellite network, we introduce the discrete-time model and the notation used throughout the remainder of this paper. We then define how eligible access satellites are selected for a UT and a GW based on visibility duration, and how these selections imply the set of ISL path candidates at each time step. This provides the foundation for the path-selection strategies defined thereafter in \Cref{sec:metrics}.

\subsection{Time Model and Notation}

We consider a discrete-time model with time steps 
\[t\in \{\,t_i = i\cdot\Delta t \mid i = 0, 1, \dots, n-1\,\},\]
where $n$ is the total number of steps, and \(\Delta t\) the duration of each step.
The ISL network is represented by the graph
\[
G = (V, E),
\]
where \(V\) is the set of satellites, \(E\) is the set of ISLs, and
\[
w_{t} : E \rightarrow \mathbb{R}_{>0}
\]
assigns a time-dependent cost to each ISL at time step \(t\).
In this work, ISL cost is defined as the Euclidean distance between neighboring satellites. 
Each satellite maintains four ISLs: two to adjacent satellites in the same orbital plane and two to satellites in adjacent planes.
For simplicity, both intra-plane and inter-plane ISL connections are assumed to remain fixed over time, since the common orbital period preserves the relative neighborhood relations between satellites. Therefore, only the link costs $w_t$ vary as satellites move along their orbits. 
Using Euclidean distance as the ISL cost provides a simple geometric baseline that can be derived directly from satellite positions without introducing additional assumptions on traffic conditions, queuing effects, link scheduling, or protocol-specific forwarding behavior. Since this metric captures only the propagation-related part of path efficiency, we additionally evaluate ISL hop count as a separate metric in Section VI.

For the subsequent path-selection model, we use the following notation:
\begin{itemize}
    \item \(\acsatu(t) \in V\): access satellite of the UT at time \(t\),
    \item \(\acsatg(t) \subseteq V\): set of eligible GW access satellites at time~\(t\), satisfying \(|\acsatg(t)| \le n_g\), where \(n_g=8\) denotes the maximum number of GW access satellites,
    \item \(\mathcal{P}(t)\): set of candidate paths from \(\acsatu(t)\) to the satellites in \(\acsatg(t)\),
    \item \(p^*(t) \in \mathcal{P}(t)\): selected ISL path at time \(t\),
    \item \( l(p,t) = \sum_{e \in p} w_t(e) \): Euclidean ISL path length of path \(p\) at time \(t\).
\end{itemize}

\subsection{Access Satellite Selection}
Satellite visiblity from a ground location is determined by a visibility cone centered on the local zenith and bounded by a minimum elevation angle above the local horizon.
A satellite is considered a potential access satellite while it is within visibility of the ground location.
Based on this principle, we select one access satellite \(\acsatu(t)\) for the UT and a set of access satellites  \(\acsatg(t)\) for the GW, based on visibility duration, as outlined in the following.
\subsubsection{Pass Direction}
A satellite is on an ascending pass when it appears to move northward relative to the Earth's surface, and on a descending pass when it appears to move southward. 
To an observer on the ground, these form two classes of visible satellites crossing the sky in opposite directions.
Although such satellites may appear close in the sky, they are generally far apart in the ISL topology. Because the considered ground positions are not located near the northernmost and southernmost extent of the constellation, this remains true throughout the evaluated scenarios.

\subsubsection{User Terminal Access Satellite}
For the UT, access satellite selection is restricted to a single satellite pass direction, either ascending or descending. This avoids oscillations between satellites belonging to different passes. 
Limiting selection to one pass direction therefore ensures a consistent UT-side starting point for candidate-path calculation.

At each time step \(t\), the UT is associated with exactly one access satellite \(\acsatu(t)\). As long as the current access satellite remains within the UT's visibility cone, it is retained. Once it leaves the cone, a new access satellite is selected from the set of satellites that are currently visible to the UT and belong to the chosen pass direction (ascending or descending). The selected satellite is the one with the longest remaining visibility interval.

\subsubsection{Gateway Access Satellites}
For the GW, instead of selecting a single access satellite, the gateway maintains a set $\acsatg(t)$ of up to $n_g$ access satellites at each time step $t$. It consists of the $n_g$ visible satellites with the longest visibility intervals at time of selection. As time evolves, satellites may leave or enter the visibility cone. Whenever a satellite in $\acsatg(t)$ leaves the visibility cone, it is removed. When $|\acsatg(t)| < n_g$, the set is replenished by adding visible satellites that have the longest remaining visibility intervals.

\subsection{Set of Path Candidates}
We compute for each gateway access satellite in  $\acsatg(t)$ the $k=10$ shortest loopless ISL paths to the UT access satellite \(\acsatu(t)\), ordered by Euclidean path length, using Yen's algorithm \cite{Ye71}, as provided by the NetworkX Python library \cite{HaSc08}.
The union of paths across all gateway access satellites forms the set of path candidates $\mathcal{P}(t)$, whose size is at most $|\acsatg(t)| \cdot k$.
A \emph{path selector} then chooses one path from the candidates based on a defined optimization goal.
Different path selectors pursue different objectives that we define in \Cref{sec:metrics}.

The choice $k=10$ ensures that each selector operates on a sufficiently rich set of alternatives while keeping candidate generation tractable.

%% file: chapters/04-metrics-selectors.tex
\section{Metrics and Path Selection}
\label{sec:metrics}
In the following section, we establish the basis for comparing the considered path-selection strategies. We first define the evaluation metrics used throughout the paper, covering both latency-related and stability-related properties. We then introduce the heuristic path selectors evaluated with respect to these metrics.

\subsection{Evaluation Metrics}
The evaluation is based on four metrics: the average of Euclidean path length, the average of ISL hop count, the path-change rate, and the used-links rate.

\subsubsection{Average Euclidean Path Length}
For each run, we compute the temporal average of the Euclidean ISL path length of the selected path:
\[
\bar{l} = \frac{1}{n}\sum_{i=0}^{n-1} l(p^*(t_i), t_i).
\]

\subsubsection{Average ISL Hop Count}
To complement Euclidean ISL path length, we also report the average number of ISL hops:
\[
\bar{h} = \frac{1}{n}\sum_{i=0}^{n-1} h(p^*(t_i)),
\]
where $h(p)$ denotes the number of ISLs contained in path $p$.

\subsubsection{Path-Change Rate}
To quantify how frequently the selected route changes over time, we define the path-change rate as
\[
r_{\mathrm{pc}} = \frac{1}{T_{\mathrm{obs}}}
\sum_{i=1}^{n-1} \mathds{1}\!\left[p^*(t_i) \neq p^*(t_{i-1})\right],
\]
where $T_{\mathrm{obs}} = (n-1) \Delta t$ is the observation time excluding the first time step and $\mathds{1}[\cdot]$ denotes the indicator function.

\subsubsection{Used-Links Rate}
Let $L(p)$ denote the set of ISLs contained in path $p$. For two consecutive paths $q$ and $p$, we define the difference in ISL
\[
\LA(q, p) = |L(p) \setminus L(q)|,
\]
that is, the number of ISLs that are used by the new path $p$ but were not used by the previous path $q$.

Summed over consecutive time steps, this captures the links that become used after the initial path and thus becomes the used-links rate
\[
r_{\mathrm{use}} = \frac{1}{T_{\mathrm{obs}}}
\sum_{i=1}^{n-1} \LA\!\left(p^*(t_{i-1}), p^*(t_i)\right),
\]
where $T_{\mathrm{obs}}$ is the observation time defined above.

\subsection{Path Selectors}
Given the candidate set $P(t)$, this section defines the three path selectors considered in the paper.

\subsubsection{Shortest-Path Selector}
As a baseline, we consider the selector that always chooses the candidate path with minimum Euclidean ISL path length. Accordingly, at time step \(t\), the selected path is given by
\[
p^*(t) = \argmin_{p \in \mathcal{P}(t)} l(p,t).
\]

The shortest-path selector only optimizes the ISL path length at the current time step. Consequently, when multiple candidates have similar Euclidean length, even minor length changes may aggressively trigger switching between paths.

\subsubsection{Maximum-Lifetime Selector}
As a stability-oriented selector, we consider a policy that prioritizes path persistence over instantaneous path cost. We select a short candidate that is expected to remain feasible for a longer time:
The selector operates as follows:
\begin{enumerate}
    \item If the previously selected path is still available, it is retained.
    \item Otherwise, among the \(j=5\) shortest candidate paths, the selector chooses the one with the longest remaining availability.
\end{enumerate}
We set $j=5$ so that the maximum-lifetime decision is made only among a limited subset of near-shortest paths.
Since UT-side access is restricted to a single pass direction, whereas GW-side access may involve both ascending and descending passes, the full candidate set may contain paths to topologically distant GW-side access satellites that are technically feasible but operationally implausible. Restricting the decision to near-shortest candidates avoids such long detours.

Formally, the selected path at time \(t\) is defined as
\begin{align*}
p^*(0) &=  \displaystyle
\argmax_{p\in\mathcal{P}_j(0)} \tau(p,0), \\[1ex]
p^*(t+\Delta t)&=
\begin{cases}
p^*(t), & \text{if } p^*(t)\in\mathcal{P}(t+\Delta t),\\[1ex]
\displaystyle
\argmax_{p\in\mathcal{P}_j(t+\Delta t)} \tau(p,t), & \text{else,}
\end{cases}
\end{align*}

where \(\tau(p,t)\) denotes the remaining availability time of path \(p\) at time \(t\), and \(\mathcal{P}_j(t)\subseteq \mathcal{P}(t)\) denotes the subset of the \(j\) shortest candidate paths.

\subsubsection{Least-Links-Used Selector}
We consider a path selector that minimizes the number of newly introduced ISLs relative to the previously selected path, as follows:
\begin{enumerate}
    \item If the previously selected path is still available, it is retained.
    \item Otherwise, the selector chooses the candidate path that minimizes the number of link additions relative to the previous path. If several candidates yield the same number of added links, the tie is resolved in favor of the path with the longest remaining availability, thereby promoting path persistence.
\end{enumerate}

Formally, the selected path at time \(t\) is defined as
\begin{align*}
p^*(0) &= \argmin_{p \in \mathcal{P}(0)} l(p,0), \\[1ex]
p^*(t+\Delta t) &=\begin{cases}
p^*(t),  \hfill \text{if } p^*(t)\in\mathcal{P}(t+\Delta t),  \\[1ex]
\displaystyle
\argminlex_{p\in\mathcal{P}(t+\Delta t)}
\left(\LA\left(p^*(t), p\right),\, \tau(p,t)\right), \qquad \text{else,}
\end{cases}
\end{align*}

where $\argminlex$ denotes the $\argmin$ taken with respect to the lexicographic order on tuples. 

%% file: chapters/05-evaluation.tex
\section{Evaluation Methodology}
\label{sec:eval}
This section outlines the simulation setup, covering both the constellation assumptions and the geographic sampling of UT and GW positions.

\subsection{Constellation Setup}

We evaluate the proposed path selectors in a simulated Walker-Delta LEO constellation,
using an Amazon-Leo-like~\cite{FCC2024Kuiper} configuration with inclination $51.9^\circ$, altitude $630\,\mathrm{km}$, and $1156$ satellites equally distributed over $34$ orbital planes. Since the phasing factor of the target system is not publicly specified, we assume a phasing factor of $17$, corresponding to half the number of orbital planes. In the absence of public design information, this serves as a neutral 
default that yields balanced staggering between adjacent orbital planes.

We use two-line element (TLE) sets to represent satellite orbits, a standardized format that describes the orbit of an Earth-orbiting object at a given epoch.
Satellite positions are computed using the Simplified General Perturbations 4 (SGP4) model, which is the standard propagator for TLE-based orbit descriptions.
Since the focus of this work is on routing behavior rather than orbit prediction accuracy, perturbation-specific parameters are not modeled beyond the nominal TLE structure.

In our implementation, SGP4 is accessed through the \textit{Skyfield} library~\cite{skyfield}, which provides positions at arbitrary times. These propagated positions are used to determine satellite visibility, derive access satellites, and compute candidate paths throughout the simulation.

\subsection{Geographic Scenario Set}
\figepsNew{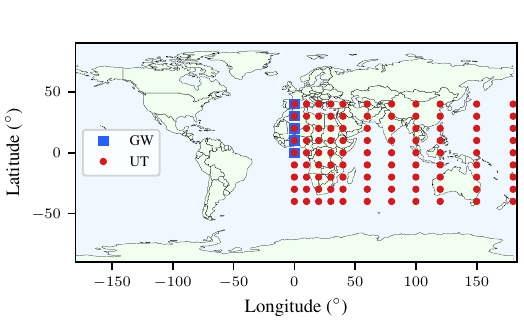}{%
Geographic placement of GWs and UTs used in the evaluation yielding $5\times 99=495$ UT--GW pairs.
}

We sample positions on a structured geographic grid to obtain a broad set of UT--GW communication scenarios, as illustrated in \Cref{fig:position_configuration_half.pdf}.
The positions are chosen such that UT and GW visibility cones remain within the interior of the constellation footprint.

Only one longitudinal half of the globe is sampled to substantially reduce the number of runs while still covering the full range of UT--GW angular separations. With that in mind, ascending and descending UT access satellite sequences are not topologically equivalent, as they correspond to different orbital pass directions and therefore yield different feasible ISL paths toward the GW-side access satellite.

The gateways with latitudes 
$0^\circ$, $10^\circ$, $20^\circ$, $30^\circ$, and $40^\circ$ are restricted to the Greenwich meridian in the northern hemisphere to keep the scenario space structured and manageable.
User terminals are placed at longitudes $0^\circ$, $10^\circ$, $20^\circ$, $30^\circ$, $40^\circ$, $60^\circ$, $80^\circ$, $100^\circ$, $120^\circ$, $150^\circ$, and $180^\circ$, and at latitudes 
$-40^\circ$, $-30^\circ$, $-20^\circ$, $-10^\circ$, $0^\circ$, $10^\circ$, $20^\circ$, $30^\circ$, and $40^\circ$. 
Larger longitude separations are sampled more sparsely to limit the number of simulation runs.
Overall, this yields $99$ UT positions and $5$ GW positions, i.e., $495$ UT--GW scenario pairs.

\subsection{Run Execution}
An observation time of $\SI{20}{\minute}$ is chosen to capture repeated path updates per scenario while keeping the simulation computationally manageable. 
Each run is discretized into 150 time steps of length $\Delta t = 8\,\mathrm{s}$.
Satellite visibility for UTs and GWs is modeled with a minimum elevation angle of $\SI{30}{\degree}$ above the local horizon to exclude low-elevation satellites.

For each UT--GW scenario pair, satellite positions, visibility relations, access satellites, candidate paths, and selected paths are updated at every time step according to the model in Sections~IV and~V.
For every retained run, we record the selected ISL path $p^*(t)$ at each time step and derive the corresponding metrics described below.
Since the UT access sequence in our setup follows either ascending or descending satellite passes, each UT--GW combination is simulated for both cases.

%% file: chapters/06-discussion.tex
\figepsFullht{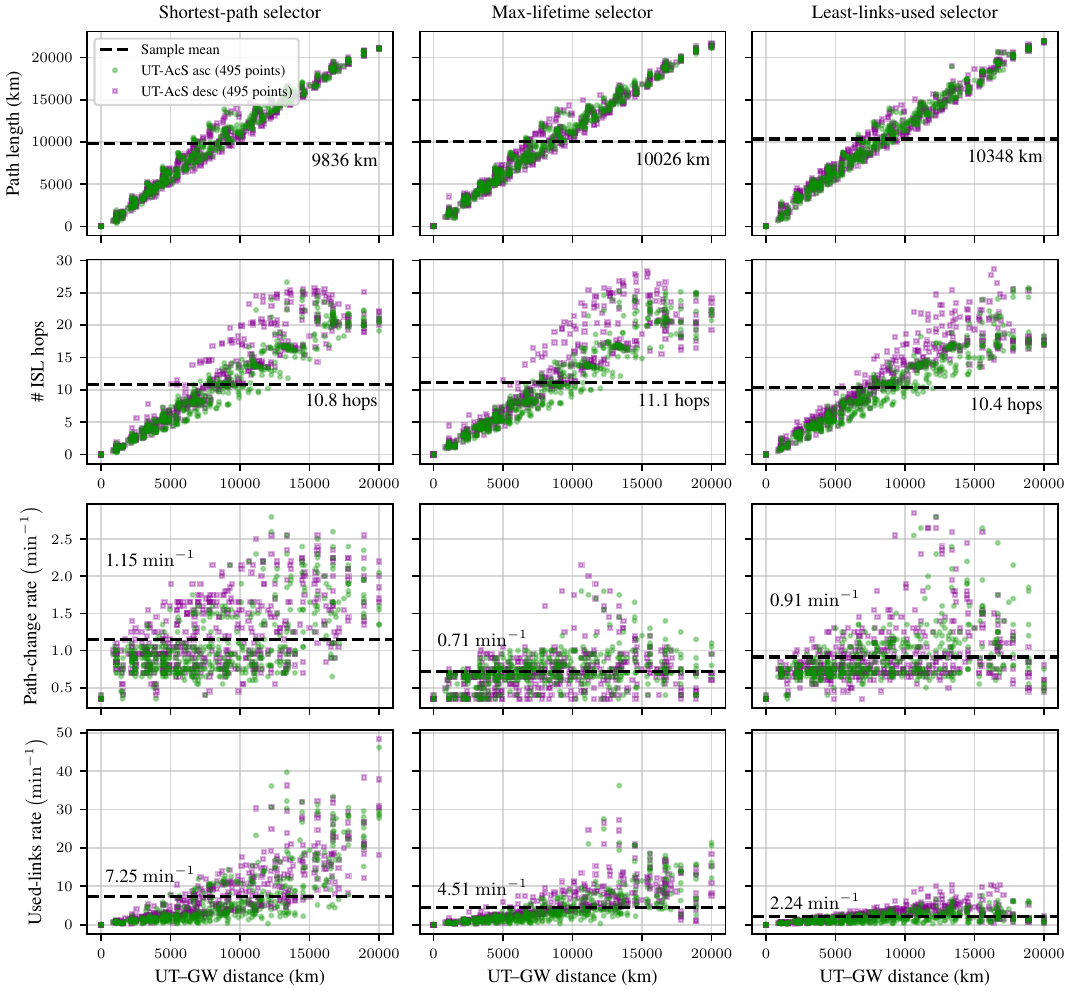}{%
We provide an overview of the evaluation results for the three considered path selectors as a function of the geodetic UT--GW distance. For each selector, the figure reports the scenario-wise values of average Euclidean path length, average ISL hop count, path-change rate, and used-links rate, separately for ascending and descending UT access sequences. The dashed horizontal line indicates the sample mean over all 495 UT--GW combinations, including both ascending and descending UT access satellite passes. The plots already reveal the central trade-off studied in this paper: the shortest-path selector yields the lowest average Euclidean path length, whereas the maximum-lifetime and least-links-used selectors reduce routing dynamics to different degrees, at the cost of moderately longer paths. In particular, the shortest-path selector achieves the lowest mean path length of \SI{9836}{\kilo\meter}, but also the highest mean path-change and used-links rates, at \SI{1.15}{\minute^{-1}} and \SI{7.25}{\minute^{-1}}, respectively. By contrast, the maximum-lifetime selector most strongly suppresses path-change, reducing it to \SI{0.71}{\minute^{-1}}, whereas the least-links-used selector most strongly limits the used-links, yielding a rate of \SI{2.24}{\minute^{-1}}.
}

\section{Results and Discussion}
\label{sec:results}
We present and discuss the evaluation results for the considered path-selection heuristics over the full set of sampled UT--GW scenarios. \Cref{fig:eval_plots_complete.pdf} reports the selector-dependent behavior of Euclidean path length, hop count, path-change rate, and used-links rate as functions of geodetic UT--GW distance. We thereby illustrate the trade-off between instantaneous path efficiency and routing stability.
Both ascending and descending UT access-satellite sequences are included, showing differences between the two cases while preserving the same overall trend.
The results show a clear pattern: the shortest-path selector achieves the smallest average Euclidean path length, but also the highest path-change and used-links rates.

By contrast, the two persistence-oriented selectors remain close to the shortest-path baseline in average Euclidean path length while substantially reducing routing dynamics. The maximum-lifetime selector is most effective in reducing path changes, whereas the least-links-used selector is most effective in limiting structural changes in the set of ISLs that carry traffic.

\subsection{Path Length and Hop Count}
Across all three selectors, the average Euclidean ISL path length increases almost linearly with geodetic UT--GW distance, as larger ground separations generally require longer ISL routes. The three selectors differ mainly in their vertical offset: the shortest-path selector attains the lowest sample mean at \SI{9836}{\kilo\meter}, the maximum-lifetime selector increases this only slightly to \SI{10026}{\kilo\meter}, and the least-links-used selector yields \SI{10348}{\kilo\meter}. Thus, the two persistence-oriented selectors incur only a modest path-length penalty relative to the shortest-path baseline.

The hop-count results contrast this observation. 
Importantly, the least-links-used selector yields the lowest average ISL hop count, as minimizing link additions implicitly favors paths with fewer hops over time.
Instead, the shortest-path selector minimizes Euclidean ISL path length and naturally does not achieve the lowest hop count. 

\subsection{Path-Change Rate}
The shortest-path selector has the highest mean path-change rate at \SI{1.15}{\minute^{-1}}, and its scatter broadens markedly at larger UT--GW distances. 
\FloatBarrier
This is consistent with the expected behavior of our shortest-path selector rule: even small changes in path length can trigger route switching.

By contrast, the path-change-rate results highlight the strongest benefit of the maximum-lifetime selector that reduces the mean path-change rate to \SI{0.71}{\minute^{-1}}, corresponding to a reduction of roughly 38\%. 
Its scatter forms a much tighter band over the full distance range. This behavior is consistent with the selector design, which reuses the current path whenever possible and otherwise prefers a near-shortest candidate with long remaining availability.

The least-links-used selector remains intermediate, with a mean path-change rate of \SI{0.91}{\minute^{-1}} and its scatter is more concentrated in the lower range than that of the shortest-path selector.

\subsection{Used-Links Rate}
The used-links-rate metric differentiates the selectors even more strongly. The shortest-path selector again performs worst, with a mean of \SI{7.25}{\minute^{-1}}, and shows both a pronounced upward trend and increasing scatter at larger distances. As expected, shortest-path updates do not merely change routes frequently; they also introduce many previously unused ISLs into the active set.

The maximum-lifetime selector reduces the mean used-links rate to \SI{4.51}{\minute^{-1}}, which is already a substantial improvement. However, the least-links-used selector performs best at \SI{2.24}{\minute^{-1}}, corresponding to a reduction of about 69\% relative to the shortest-path selector. Its scatter remains tightly concentrated near low values over almost the entire distance range, indicating that its objective translates directly into strong structural stability of the active ISL set.

\subsection{Scope and Limitations}
This evaluation provides a controlled comparison of path-selection strategies, not a full operational model of a deployed LEO network. Euclidean ISL path length is used as a simple geometry-based proxy for propagation-related path efficiency, but it omits important delay components such as processing, scheduling, queuing, and forwarding delays. Since these components may scale more strongly with the number of traversed ISLs than with geometric distance, we evaluate ISL hop count separately. The structured UT--GW grid enables comparable scenario coverage, but does not reflect sparse and non-uniform real deployments. Moreover, the used-links rate captures structural path changes, but not actual link utilization under traffic demand. Future work should evaluate these effects under realistic gateway and user terminal distributions and extend the model to include the omitted delay components and traffic-dependent link utilization.

%% file: chapters/07-conclusion.tex
\section{Conclusion}
\label{sec:conclusion}

Inter-satellite paths between a user terminal and a gateway change about every minute.
We considered several strategies for selecting a replacement path from a set of candidates. 
We studied how they impact the following metrics: path length in terms of Euclidean distance and number of hops, path change rate, and rate of used links. They are relevant as they either add latency to communication or complexity to control and resource management. 

We explained how candidate paths are chosen and defined three path selection heuristics that minimize path lengths, change rate, and rate of used links: the shortest-path selector, the maximum-lifetime selector, and the least-links-used selector. 
Our results show that the heuristics indeed achieve the best values for the metrics they were designed for.
Moreover, the shortest-path selector reduces path length only by about 5\% compared to the other methods. 
The results are averaged over 495 different user-terminal and gateway positions, covering distances from zero to \SI{20000}{\kilo\meter}.

In contrast, a least path change rate of 0.71 changes/min could be achieved for the maximum-lifetime selector while the shortest-path selector leads to 1.15 changes/min and the least-links-used selector to 0.91 changes/min. The rate of used links could be effectively minimized to 2.24 new links per minute by the specialized selector while the shortest-path selector and the maximum-lifetime selector required 7.25 and 4.51 new links per minute, respectively. 

Since the differences are significant, these findings should be considered in the design and operation of routing in LEO satellite constellations.
Future work may investigate multi-objective selectors that jointly optimize latency, path persistence, and link reuse.